# Incorporating Personality Traits in User Modeling for EUD

Federica Cena[1], Cristina Gena[1], Claudio Mattutino[1], Michele Mioli[1], Barbara Treccani[2], Fabiana Vernero[1,*] and Massimo Zancanaro[2]

[1]*University of Turin, Department of Computer Science, C.so Svizzera 185, 10149 Turin, Italy*
[2]*University of Trento, Department of Psychology and Cognitive Sciences, Corso Bettini, 31, 38068 Rovereto. Italy*

**Abstract**

Personality traits such as Need for Cognition, Locus of Control, Mindset and Self-efficacy could impact the perception, acceptance and appreciation of recommendations provided to support configuration tasks in the End User Development (EUD) context. In this paper we describe the user model services we have developed to measure such traits. These services can be accessed by users through a simple web interface and can be queried by EUD applications by means of REST API.

## 1. Introduction

While smart environments can be understood as ecosystems of interacting objects, such as sensors, devices, appliances and embedded systems [1] which seamlessly cooperate in order to help users carry out their tasks [2] and, ultimately, improve their lives [3], end-users are left with the daunting task of configuring the environment itself. In fact, not only does configuration imply making *preferential choices* [4] on a number of different aspects (e.g., what IoT components should be included, what events should trigger intelligent behaviours), but end users might also lack the technical knowledge which could help them better understand and, therefore, control their IoT environment.

Recommender systems can support users' decision making processes [5] and user preferential choices [4] by providing personalized and non-personalized [6] suggestions. More specifically, it has long been demonstrated that recommendations can improve system usability and user experience in the End-User Development (EUD) domain [7], where end-users are required to act as "non-professional software developers" [8] who can create, modify, or extend software artifacts. For example, Haines et al. [7] put forward that recommendations for end users could regard shared procedures, preferred defaults and examples, among other things.

Starting from these observations, in our past work carried out in the context of the EMPATHY project[1] we experimented with personalized recommendations as a way to help users make configuration choices [9]. We assumed that end-users have a certain degree of freedom on the choice of the smart objects to include in their ecosystem, and that configuration tasks basically consist in the definition of trigger-action rules such as the following, meant for a smart home context: *if a weather station (trigger object) measures that the level of humidity in the air is above a certain threshold (trigger event), the automatic irrigation system (action object) is disabled (action event)*. The recommendation services we built suggest smart objects to couple with an input object chosen by the end-user, with the aim of composing a trigger-action rule. More specifically, possible *action* objects are recommended to end-users who start their rule by specifying a trigger, while *trigger* objects are suggested to end-users which input an *action* object.

Nevertheless, while recommendations can be broadly considered useful for any end-user, their actual perception, acceptance and appreciation can depend on individual features. For example, [10] showed that personality plays an important role in determining user preferences and decision-making strategies. Personality traits such as Need for Cognition, Locus of Control and Mindset, which are linked to the area of decision-making, were found to have an effect on user behaviour within a recommender system [11, 12, 13], while the related trait of Self-efficacy can impact general technology acceptance [14]. To the best of the authors' knowledge, however, the relation between users' psychological traits and recommendations is still unexplored in the EUD domain and in connection with configuration tasks.

Following on from these results, we put forward that incorporating information on such personality traits in user models can help generate more effective recommendations for endusers. For example, the quantity of recommendations as well as the presence and level of detail of explanations could be tailored according to user personality. As part of our research within the EMPATHY project, we are carrying out a series of experiments aimed at determining the impact of personality traits on recommendations in the specific context of a configuration task, an aspect which has not yet been explored in relevant literature. In addition, we have developed a few web services which allow to measure different personality traits, making use of standard scales, and which are publicly made available to other researchers.

In this paper, after having discussed related work (Section n2), we present our web services (Section 3), which are part of a platform we built to experiment with recommendations (Section 4). Conclusions (Section 5) complete the paper.

---

[1] http://www.empathy-project.eu/

## 2. Related work

**Personality in recommender systems.** Personality explains individual differences in emotional, interpersonal, experiential, attitudinal and motivational styles [15]. Personality-related information has been used in recommender systems to help overcome various issues, such as the *cold start* problem [16] or satisfaction with the suggested options in a group recommendation scenario [17]. While most work has referred to the Big-Five model [15], which provides an overall picture of personality in terms of five broad dimensions, some studies have focused on specific traits. For example, Mindset was found to influence users' evaluation of recommendations at various levels: perceived effectiveness [18], likelihood to be influenced [19] and acceptance [13]. Need for Cognition has an impact on user satisfaction with explanations [11] and willingness to rely on recommendations [20]. Locus of Control has an effect on users' tendency to trust recommendations [12], while, to the best of the authors' knowledge, *Self-efficacy* was not studied in connection with recommender systems, but significantly affects users' acceptance of different technologies [14, 21].

**Recommendation in the EUD domain** Among the earliest examples of recommendations in the EUD domain are systems such EAGER, Dynamic Macro, and APE [22, 23, 24], which adopt a programming by example (PBE) approach to learn how to complete users' tasks based on the observation of their behaviour and therefore recommended automation directly within the user's workflow. In a similar vein, Lumière [25] and WARP [26] used probabilistic models for activity recognition to offer context-dependent assistance. While Lumière provided suggestions on a limited number of predefined tasks series of predefined tasks, WARP was also able to continuously extend its knowledge. Task Assistant also generated recommendations over an extensible knowledge base, exploiting manual procedures defined through EUD [27]. Nevertheless, Haines et al. [7] state that only a limited number of systems currently adopt recommendations to support decision-making within EUD, and suggest that recommendations could prove useful in four scenarios: 1. inserting automation into the user's workflow; 2. helping the user make the right decisions; 3. handling errors; 4. supporting unplanned sharing.

## 3. User model services

With the aim of fostering research on the interplay between personality and recommendations in the EUD domain, as well as offering a public service to researchers and practitioners who wish to incorporate information on personality in their recommender systems, we have built a mechanism to calculate four different, but closely related traits, designing a specific test for each of them:

1. **Self efficacy** (https://app.empathy.di.unito.it/#/selfEfficacy). This construct refers to individuals' beliefs about their ability to exercise control over their own functioning and activities [28]. To measure it, we used the the 10-item Self-efficacy IPIP scale[2].

---

[2] https://ipip.ori.org/newCPIKey.htm#Self-Efficacy)

2. **Need for cognition** (https://app.empathy.di.unito.it/#/needForCognition). This construct describes individuals' tendency to engage in and enjoy thinking and, in general, cognitively demanding activities[29]. To measure it, we used the 10-item IPIP scale[3].

3. **Locus of control** (https://app.empathy.di.unito.it/#/locusOfControl). This construct represents whether individuals believe they have weak (external) or strong (internal) control over the events that affect their lives [30]. To measure it, we used the 5-item IPIP rational scale[4]

4. **Mindset** (https://app.empathy.di.unito.it/#/mindset). This construct refers to implicit theories which create a sort of framework and then stimulate judgments and reactions which are coherent with that framework. To measure it, we formulated a set of questions inspired by the Growth Mindset Scale by Dweck et al. [31], which investigates the beliefs in fixed versus malleable human attributes. Differnetly from the original scale, we focused on problem solving instead of intelligence.

Each test is designed to provide a mechanism to profile their users to external applications. On the one hand, users can perform one or multiple tests, according to the application needs, accessing the aforelisted web pages. On the other hand, the prototype also exposes two REST API which provide the test results to querying applications in order to promote interoperable user modeling exchanges [32], Berkovsky et al. [33]. Results for a specific user can be retrieved by providing their username. Thresholds ("low", "medium", "high") for the four traits were calculated based on the dataset collected in a previous experiment [34].

In particular, one API provides an overview of the personality traits for the selected user: https://app.empathy.di.unito.it/api/empathy/userPsychometrics/[username]. A response example follows:

```
{
    "locusOfControl": "external",
    "needForCognition": "low",
    "selfEfficacy": "low",
    "username": "Donald.Duck"
}
```

The other API provides more details on the traits by returning numerical values: https://app.empathy.di.unito.it/api/empathy/userPsychometricsValues/[username]. A response example follows:

```
{
    "locusOfControl": 1.8,
```

---

[3] https://ipip.ori.org/newSingleConstructsKey.htm#Need-for-Cognition
[4] https://ipip.ori.org/newSingleConstructsKey.htm#Locus-of-Control

```
    "rangeLocusOfControl": "2.885 - 3.615",
    "needForCognition": 3,
    "rangeNeedForCognition": "3.46 - 3.98",
    "selfEfficacy": 3,
    "rangeSelfEfficacy": "3.52 - 3.96",
    "username": "Donald.Duck"
}
```

## 4. Prototype recommendation platform

In the context of the EMPATHY project, we have built a prototype platform[5] to experiment with recommendations and connected services. Apart from the psychometric user model services described in Section 3, the platform hosts:

- A **user modeling component** (under development) which includes individual characteristics and psychological traits as well as owned smart objects, user preferences for smart objects and smart home goals (such as energy saving, safety, comfort, etc.).

- A **testbed for similarity algorithms**, where the researcher can choose both the preferred measure (Jaccard, Pearson, Cosine, Simple matching) and the user model features to include.

- A **series of recommendation services** which suggest smart objects (either a *trigger* or an *action* object) to couple with an input object chosen by the end-user. A web interface where these services can be accessed is also available: more specifically, one page suggests *action* object categories, given a *trigger* object category, while the other one suggests the *trigger* object categories, given an *action* object category. Example rules are also suggested for each trigger-action association. Knowledge on suitable trigger-action associations was derived by applying the *association rules* technique (Apriori algorithm) on an IFTTT[6] rules dataset.

**Implementation details.** The web interface is developed using Vue.js and Vuetify as a material design framework. The application server is implemented with Spring Boot and exposes a set of REST API. The recommender is a module built in Java which contains all the logic required to provide the different recommendation techniques. The data are stored in a mongoDB instance and are organized in different collections and databases.

## 5. Conclusion

In this paper, we have presented four web services which calculate the personality traits of Need for Cognition, Locus of Control, Mindset and Self-efficacy, which we believe can impact users' behaviour in recommender systems in the context of configuration tasks. As future work, we are planning to provide guidelines on the use of personality information to tailor recommendations in the EUD domain, based on the results of the experiments we are currently carrying out.

---

[5] https://app.empathy.di.unito.it

[6] If This Then That (IFTTT) is a private commercial company that runs services that allow a user to program a response to events (https://ifttt.com/).

## Acknowledgments

This work is partially supported by the Italian Ministry of University and Research (MIUR) under grant PRIN 2017 "EMPATHY: EMpowering People in deAling with internet of THings ecosYstems".